\documentclass[aps,pra,reprint,footinbib,footinbib,floatfix,superscriptaddress]{revtex4-1}
\usepackage{amsmath}
\usepackage{amssymb}
\usepackage{graphicx} 

\def\be{\begin{equation}}
\def\ee{\end{equation}}
\def\ber{\begin{eqnarray}}
\def\eer{\end{eqnarray}}
\def\bern{\begin{eqnarray*}}
\def\eern{\end{eqnarray*}}

\def\rv{\mathbf{r}}

\def\Gv{\mathbf{G}}

\def\pv{\mathbf{p}}
\def\qv{\mathbf{q}}

\def\bv{\mathbf{b}}

\def\ev{\mathbf{e}}
\def\Fv{\mathbf{F}}

\def\Rv{\mathbf{R}}

\def\uv{\mathbf{u}}

\def\0v{\mathbf{0}}
\def\1v{\mathbf{1}}
\def\2v{\mathbf{2}}
\def\3v{\mathbf{3}}

\def\zv{\mathbf{z}}

\def\Yv{\mathbf{Y}}

\def\pa{\partial}

\begin{document}

\title{Coupled plasmon - phonon excitations in extrinsic  monolayer graphene}

\author {Vladimir U. Nazarov}
\affiliation{Research Center for Applied Sciences, Academia Sinica, Taipei 11529, Taiwan}
\affiliation{Qatar Environment and Energy Research Institute, Qatar Foundation, Doha, Qatar}
\email{nazarov@gate.sinica.edu}

\author{Fahhad  Alharbi}
\affiliation{King Abdulaziz City for Science and Technology, Riyadh, Saudi Arabia}
\affiliation{Qatar Environment and Energy Research Institute, Qatar Foundation, Doha, Qatar}

\author{Timothy S. Fisher}
\affiliation{Birck Nanotechnology Center and School of Mechanical Engineering, Purdue University, West Lafayette, IN 47906, USA}
\affiliation{Qatar Environment and Energy Research Institute, Qatar Foundation, Doha, Qatar}

\author{Sabre Kais}
\affiliation{Department of Chemistry,  Physics and Birck Nanotechnology Center, Purdue University, West Lafayette, IN 47907 USA}
\affiliation{Qatar Environment and Energy Research Institute, Qatar Foundation, Doha, Qatar}

\begin{abstract}
 The existence of an acoustic plasmon in extrinsic (doped or gated) monolayer graphene  was found recently in an {\it ab initio} calculation with the frozen lattice
[M. Pisarra {\it et al.}, arXiv:1306.6273, 2013].
By the  {\em fully dynamic} density-functional perturbation theory approach,
we demonstrate a strong coupling of the acoustic plasmonic mode to  lattice vibrations.
Thereby, the acoustic plasmon in graphene  does not exist as an isolated 
excitation, but it is rather bound into a combined plasmon-phonon mode.  
We show that the coupling provides a mechanism for the {\em bidirectional} 
energy exchange between the electronic and the ionic subsystems with fundamentally, as well as
practically, important implications for  the lattice cooling and heating by electrons in graphene. 
\end{abstract}


\maketitle

Known for its extraordinary properties and vast potential applications \cite{Neto-09},
graphene -- a two-dimensional crystal comprised of a honeycomb lattice of carbon atoms --
continues to receive much attention  as it reveals new remarkable features
\cite{Park-09,Kogan-12,Kogan-13,Nazarov-13,Kang-13,Pisarra-13}. 
For one of the recent findings, an acoustic plasmon (APl) (plasmon with {\em linear}
wave-vector dispersion)
has been  predicted theoretically 
in an extrinsic free-standing monolayer graphene \cite{Pisarra-13}. This finding is extraordinary considering
that APl generation conventionally involves 
a surface state immersed in the bulk of a metal \cite{Pitarke-04}.

Exhibiting linear wave-vector dispersion, acoustic APl persists down to low frequencies, 
where it can be expected to interact with phonon oscillations.
The possibility of coupling these two types of elementary excitations
motivates questions  of fundamental physics as well as of potential applications. 
In this Letter we show that the APl -  phonon coupling  indeed occurs 
in the electron-doped graphene and it provides a mechanism for the 
bidirectional energy exchange between the electronic and ionic subsystems.
The conventional treatment of lattice vibrations by  
frequency-independent density-functional perturbation theory (DFPT) \cite{Giannozzi-05}
is inadequate for capturing the  essentially dynamic nature of the coupled plasmon-phonon modes, and we
therefore implement a fully {\em dynamic} approach treating 
the electron-hole, plasmon, and phonon
elementary excitations on the equal footing \cite{Bostwick-07}. 

Our {\em ab initio} calculations for  monolayer graphene  employ 
the full-potential linear augmented plane-wave (FP-LAPW)  code Elk \cite{Elk}.
The super-cell geometry is utilized with a separation of the layers in the $z$ direction  of 40 bohr, 
which effectively ensures the non-interaction between the layers.
The local-density approximation to the exchange-correlation potential  \cite{Dirac-30,Perdew-92} is used. 

\begin{figure} [h!] 
\includegraphics[width=1.0 \columnwidth,clip=true, trim=35 0 10 0]{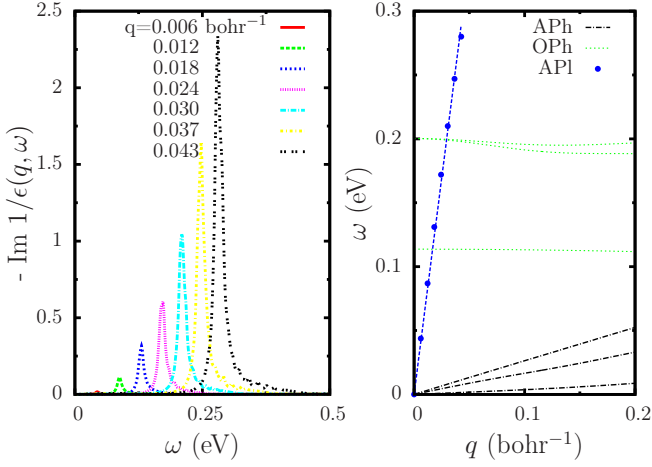} 
\caption{\label{plph} (color online) 
Left: Energy-loss function of the monolayer graphene doped with 
$\frac{1}{16}$ 
 electrons per unit cell ($1.19\times 10^{14}$ cm$^{-2}$).  
Plasmon peaks with linear (acoustic) dispersion are dominant in the 
low-frequency range of the spectra. The direction of the wave-vector $q$ is along the primitive reciprocal lattice vector.
Right: Phonons (acoustic, black dash-dotted lines, optical, green dotted lines, respectively) and acoustic plasmon (blue  symbols) dispersion. Blue dashed line is the linear best fit to the acoustic plasmon dispersion.
}
\end{figure}

{\em Acoustic-plasmon and phonons in graphene.--}
We start by reproducing the APl and phonon spectra of graphene without the coupling of the two excitations.
In Fig.~\ref{plph}, left panel, the energy-loss function of graphene 
is plotted for a number of equidistant values of the wave-vector.
The calculation with the carbon atoms fixed at their equilibrium positions has been used.
The APl can be easily recognized by the linear dispersion of the peak with the wave-vector,
which is in agreement with the recent findings in extrinsic graphene 
obtained with the use of the pseudopotential method \cite{Pisarra-13}. 
In Fig.~\ref{plph}, right panel, we plot the phonon dispersion spectra in graphene together with 
the APl dispersion derived from the energy-loss function. Acoustic plasmonic and optical phononic (OPh) dispersion curves 
intersect, which fact suggests their interaction and 
constitutes the main motivation of the subsequent study of the coupled modes. 

{\em Coupled plasmon-phonon modes.--} 
We treat the problem of coupled plasmon-phonon oscillations within the {\em dynamic} (frequency-dependent)
linear-response theory: Self-consistently,
ions are driven by an externally applied AC electric field and by the Coulomb field of moving
electrons, and in turn, electrons move under the action of the external field and the field of moving ions. 
We  consider an infinite two-dimensional (2D) crystal lying in the $xy$ plane. 
The 2D lattice vectors are denoted by $\Rv$ while the position
of the $\alpha$-th atom within the unit cell is $\bv_\alpha$. A weak external
potential of the form
\begin{equation}
\delta \phi^\textit{ext}(\rv,t)=\delta \phi^\textit{ext}(\qv,z,\omega) e^{i (\qv\cdot\rv-\omega t)}
\end{equation}
is applied to the system, where $\rv$ is the 3D position coordinate vector and $\qv$ is the 2D wave-vector. 
We seek the response {\em including
the ionic oscillations} around their equilibrium positions with the
displacements given by
\begin{equation}
\uv_{\alpha\Rv}(t)= \uv_{\alpha\Rv}(\omega)e^{- i \omega t} =\ev_\alpha e^{i (\qv\cdot\Rv-\omega t)} 
\label{disp}
\end{equation}
with 3D vectors $\ev_\alpha$.
The total Coulomb potential in the system is
\begin{equation}
\phi(\rv,t)=\phi_0(\rv)+\delta \phi(\rv) e^{-i \omega t},
\end{equation}
where
$\phi_0$ is the ground-state Coulomb potential and $\delta \phi$ is its first-order perturbation.
The force experienced  by the $\alpha$-th ion in the $\Rv$-th unit cell is
\begin{equation}
\Fv_{\alpha\Rv}(t) = - \left. Z_\alpha \nabla \phi^\textit{eff}_{\alpha\Rv}(\rv,t)\right|_{\rv=\bv_\alpha+\uv_{\alpha\Rv}(t)+\Rv},
\label{F}
\end{equation} 
where $Z_\alpha$ is the charge of the $\alpha$-th ion within the unit cell and $\phi^\textit{eff}_{\alpha\Rv}$ is the total Coulomb potential minus the self-interaction of the $(\alpha\Rv)$-th ion
\begin{equation}
\phi^\textit{eff}_{\alpha\Rv}(\rv,t)=\phi(\rv,t)-\frac{Z_\alpha}{|\rv-\bv_\alpha-\uv_{\alpha\Rv}(t)-\Rv|}.
\label{eff}
\end{equation}
Expansion of Eq.~(\ref{eff}) to the first order in the perturbation gives
\begin{equation}
\begin{split}
\phi^\textit{eff}_{\alpha\Rv}(\rv,t) & =\phi_0(\rv)-\frac{Z_\alpha}{|\rv-\bv_\alpha-\Rv|}+\delta \phi(\rv) e^{-i \omega t}\\
&+ \uv_{\alpha\Rv}(t)\cdot \nabla \frac{Z_\alpha}{|\rv-\bv_\alpha-\Rv|}.
\end{split}
\label{b1}
\end{equation}
Since
\begin{equation}
\phi_0(\rv)= v^\textit{ext}(\rv) -
\int \frac{n_0(\rv')}{|\rv-\rv'|} d\rv',
\label{b2}
\end{equation}
where $n_0(\rv)$ is the ground-state electron particle-density
and $v^\textit{ext}(\rv)$ is the equilibrium ions' potential
\begin{equation}
v^\textit{ext}(\rv)=
\sum\limits_{\beta\Rv} \frac{Z_\beta}{|\rv-\bv_\beta-\Rv|},
\label{b3}
\end{equation}
we can write for the force acting on the $\alpha$-th ion in the $\0v$-th cell
\begin{equation}
\begin{split}
\Fv_{\alpha}  & =  -  Z_\alpha \left\{  
(\ev_{\alpha}  \cdot  \nabla) \nabla \left[ \sum\limits_{(\beta\Rv)\ne(\alpha\0v)} \frac{Z_\beta}{|\rv  -  \bv_\beta  -  \Rv|} \right. \right. \\ & \left. \left. 
  -   \int  \frac{ n_0(\rv')}{|\rv-\rv'|} d\rv' \right]
  +  
\nabla  \delta \phi^\textit{eff}_{\alpha\0v}(\rv,\omega) 
\right\}_{\rv=\bv_\alpha} ,
\end{split}
\label{FFF}
\end{equation} 
where the corresponding 0-th order term has been set to zero because  ions are in their equilibrium positions  in the crystal's ground-state. 
The electronic response is governed by the equation
\begin{equation}
\delta \phi^\textit{ext}(\rv,t) \! + \! \delta \phi^I_b(\rv,t) \! = \! 
\! \! \int \! \! \epsilon(\rv,\rv',t \! - \! t') \delta \phi(\rv',t') d\rv' d t',
\label{res}
\end{equation}
where
\begin{equation}
\delta \phi^I_{b}(\rv,t)=-\sum\limits_{\alpha\Rv}   \uv_{\alpha\Rv}(t) \cdot \nabla \frac{Z_\alpha}{|\rv-\bv_\alpha-\Rv|}
\label{bI}  
\end{equation}
is the ionic displacement bare potential and $\epsilon$ is the nonlocal dielectric function of the ideal crystal. 

Based on Eqs.~(\ref{F}) and (\ref{b1}) - (\ref{bI}), 
a rather lengthy algebra, which we have included in  the Appendix,
leads to the following expression for the force
\begin{equation}
\begin{split}
F_{\alpha i}  & = F^\textit{ext}_{\alpha i}(\qv,\omega)  +  F^\textit{ei}_{\alpha i}(\qv,\omega) -\sum\limits_{\beta k}   D_{\alpha i,\beta k} (\qv) e_{\beta k} \\
& +   \sum\limits_{\beta k}  
\left[ Q_{\alpha i,\beta k} (\qv,\omega) -Q_{\alpha i,\beta k} (\qv,0) \right] e_{\beta k},
\end{split}
\label{forceD}
\end{equation}
where $D_{\alpha i,\beta k}$ are the so called dynamic matrices 
\footnote{In the context of this work, the conventional term
{\em dynamic matrices} contains ambiguity  since, in fact, they account exactly for the {\em static}
(frequency independent) part of the force acting on an ion.}
of the conventional DFPT \cite{Giannozzi-05} and
\begin{equation}
\begin{split}
 F^\textit{ext}_{\alpha i} & =
-Z_\alpha \sum\limits_{\Gv} e^{i(\Gv+\qv)\cdot\bv_\alpha} \\
& \left.  \times
\left[ i(G_i+q_i) +\hat{z}_i \frac{\pa}{\pa z}  \right] \phi^\textit{ext}(\Gv+\qv,z,\omega) \right|_{z=0}, 
\end{split}
\end{equation}
\begin{equation}
\begin{split}
F^\textit{ei}_{\alpha i} \! & = \!
  2\pi Z_\alpha \! \! \sum\limits_{\Gv\Gv'}  
\int \! 
Y_i(z,\Gv+\qv) \chi_{\Gv\Gv'}(\qv,z,z',\omega)  \\
& \times
e^{i(\Gv  +  \qv)\cdot\bv_\alpha}
\phi^\textit{ext}(\Gv' \! + \! \qv,z',\omega) d z d z',
\end{split}
\end{equation}
\begin{equation}
\begin{split}
&Q_{\alpha i,\beta k} (\qv,\omega) =
\frac{(2\pi)^2}{s_0} \sum\limits_{ \Gv\Gv'} Z_\alpha Z_\beta e^{i(\Gv+\qv)\cdot\bv_\alpha} e^{-i(\Gv'  +  \qv)  \cdot  \bv_\beta} \\
& \times \int Y_i(z,\Gv+\qv)    \chi_{\Gv\Gv'}(\qv,z,z',\omega) 
  Y_k(z',\Gv'+\qv) d z d z',
\end{split}
\end{equation}
where $\chi_{\Gv\Gv'}$ is the interacting-particles density-response function of the ideal crystal, 
$s_0$ is the area of the unit cell,
\begin{equation}
\Yv(z,\pv)=e^{-p |z|} \left[ \hat{\zv} \, sgn(z) - i \frac{\pv}{p} \right],
\label{Y}
\end{equation}
and $\hat{\zv}$ is the unit vector in the $z$ direction.

In Eq.~(\ref{forceD}), the first two terms are due to the dynamically screened external force  in the ideal crystal and
the third term is the statically screened restoring force of the displacement of the ions. The fourth term contains all the effects responsible for dynamic electron-phonon interaction.
Obviously, with the neglect of the latter ($\omega=0$ in the fourth term), Eq.~(\ref{forceD})
reduces to the conventional static DFPT case \cite{Giannozzi-05}. 

With the use of Eq.~(\ref{disp}), Newton's second law gives for the $\alpha$-th nucleus
at the $\0v$-th unit cell
\begin{equation}
-M_\alpha \omega^2 \ev_\alpha = \Fv_{\alpha}.
\label{Nl}
\end{equation}
Equations (\ref{forceD}) - (\ref{Nl}) form a $3 N $ system of linear equations for $3 N$ unknowns $e_{\alpha,i}$, $i=1,2,3$,
where $N$ is the number of atoms in an elementary unit cell.

Energy absorbed by the unit cell of the lattice per unit time is
\begin{equation}
W=\frac{\omega}{2} \sum\limits_\alpha {\rm Im} \, \left( \Fv^\textit{ext}_\alpha \cdot \ev_\alpha^* \right) .
\end{equation}

{\it Calculations and results.--}
We excite the system with the external potential
\begin{equation}
\phi^\textit{ext}(\rv,t)=e^{q z} e^{i(\qv\cdot\rv-\omega t)}
\end{equation}
and solve for the amplitudes of the oscillations $\ev_\alpha$ using the above formalism
with the random-phase approximation to the density-response matrix $\chi_{\Gv\Gv'}$.

\begin{figure}[h] 
\includegraphics[width=1.0 \columnwidth, clip=true, trim =35 0 25 0]{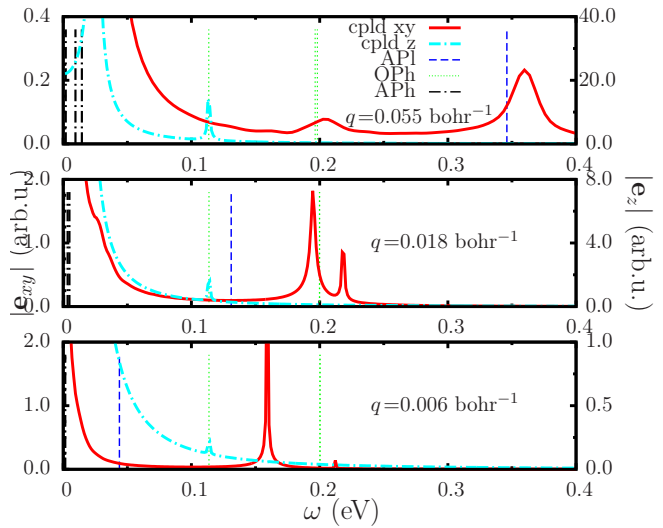} 
\caption{\label{exyz} (color online) 
Amplitude of the ions' oscillations as a function of the frequency of the applied field.
Red solid line and cyan dash-dotted line are the coupled phonon-plasmon oscillations
with $xy$- and $z$-polarization, respectively. 
The green dotted and black dash-dotted vertical lines
show the positions of the optical and acoustic phonons, respectively, while
the blue vertical dashed lines are the positions of the maxima of acoustic plasmon calculated with the {\em frozen} lattice.
}
\end{figure}

In Fig.~\ref{exyz}, the amplitude of ionic oscillations is plotted as a function of the frequency
for three values of the wave-vector. The $q$-vector dependence  
of the $xy$-polarized coupled excitation is strongly
influenced by that of the APl and $xy$-polarized OPh,
while the former is very different from the both latter.
Two coupled modes originated from the $xy$-polarized OPh and APl are prominent.
First of them has a strong $q$-dispersion, 
while the second is bound in the vicinity of $xy$-polarized OPh.
Both modes are blue-shifted compared with the APl and OPh, respectively. 
For larger $q$ (upper panel of Fig.~\ref{exyz}), the frequency of APl becomes too high
for ions to follow the oscillations, leading to the coupled mode  convergence to the APl.

The $z$-polarized mode remains practically non-dispersive and, acquiring a finite but small
line-width,  is pinned at the position of the corresponding OPh.

\begin{figure}[h] 
\includegraphics[width=1.0 \columnwidth, clip=true, trim =30 0 10 0]{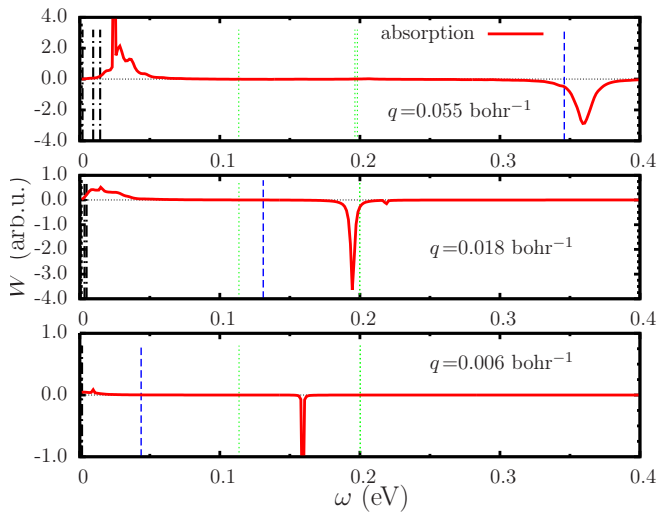} 
\caption{\label{W} (color online) 
Energy absorption by the unit cell of the lattice per unit time (red solid line).
The position of optical phonons
are shown with green dotted lines, 
while the blue dashed lines are the positions of the maxima of acoustic-plasmon 
in the calculation with the frozen lattice.}
\end{figure}

In Fig.~\ref{W}, we plot the energy absorbed by the unit cell of the lattice per unit time.
The remarkable feature in this figure is that, depending on the frequency range,
the lattice either receives the energy (positive $W$) or gives it away to the electronic
subsystem (negative $W$).
We anticipate that this phenomenon will be experimentally observable in two-terminal suspended graphene experiments. For example, 
Yi\ifmmode \breve{g}\else \u{g}\fi{}en {\it et al.} \cite{Yigen-13} recently demonstrated the ability to distinguish electronic from phononic heat conduction in a self-heated suspended device. The associated analysis of electron-phonon scattering did not, however, include plasmonic effects, which would be observable at moderate temperatures and under AC fields near the resonances predicted here.
It must be also noted that thorough understanding of APl-phonons interactions
is particularly important in the field  of superconductivity \cite{Ganguly-72}.

\begin{figure}[h] 
\includegraphics[width=1.0 \columnwidth, clip=true, trim =30 0 5 0]{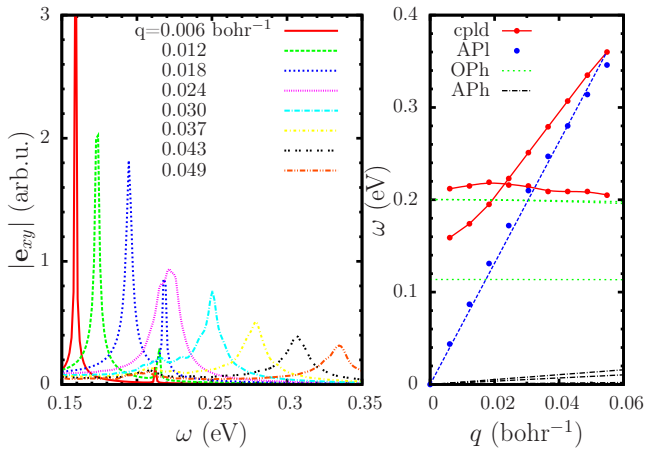} 
\caption{\label{cpldisp} (color online) 
Dispersion of the coupled  mode with the $xy$ polarization
represented by the amplitude of an ion oscillation vs. the frequency $\omega$ at a number of
the wave-vector $q$ values (left) and the dispersion law $\omega(q)$ (red solid lines) derived
from the plots in the left panel (right). Dispersion of the optical  and acoustic phonons
is shown with the green dotted and black dash-dotted lines, respectively.}
\end{figure}

Figure~\ref{cpldisp}, right panel, shows the $q$-vector dispersion derived
from the ions' oscillations amplitude dependence on the frequency (left panel).
We conclude, that at smaller $q$-vectors, the acoustic-like linear dispersion of
the coupled mode is lost, indicating that in this regime APl in graphene is an artefact of the frozen-lattice approximation.
The second branch of the coupled mode remains close and above the $xy$-polarized OPh, varying non-monotonically and eventually
converging to the latter.

In conclusion, we have implemented a fully dynamic (frequency-dependent density-functional perturbation theory)
calculation of  coupled electron-lattice oscillations in graphene.
The coupled mode behaves quite differently from the individual phonon
and acoustic plasmon modes, previously known in graphene, and the former replaces the two latter,
as acoustic plasmons and phonons do not exist in graphene by themselves, but they constitute a unified excitation of the electronic and ionic subsystems.
The coupling provides a mechanism for the transfer of energy
between the electronic subsystem and the lattice, which is shown to go in both 
directions depending on the frequency range. 
From this, promising pathways of tunable heating and cooling of the lattice by the electronic subsystem can be clearly previewed.

\acknowledgments
V.U.N. acknowledges the support from National Science Council, Taiwan,
Grant No. 100-2112-M-001-025-MY3.
T.S.F. acknowledges the support of the US Office of Naval Research (award \# N000141211006, PM: Dr. Mark Spector). 
V.U.N. and T.S.F. are
grateful for the hospitality of Qatar Environment and Energy Research Institute, Qatar Foundation, Qatar. 


%


\onecolumngrid

\appendix*
\section{Derivation of Eq.~(\ref{forceD})}

We perform  the 2D Fourier transform of Eq.~(\ref{bI})
\begin{equation}
\begin{split}
\delta \phi^I_{b}(\Gv  \! +  \! \qv,z,\omega) \!  = \! \frac{2\pi }{s_0}   \!
\sum\limits_\alpha  \! Z_\alpha  \ev_\alpha \! \cdot \! \Yv(z,\Gv \! + \! \qv) e^{-i(\Gv \! + \! \qv)   \cdot  \bv_\alpha},
\end{split}
\end{equation}
where the vector function $\Yv$ is defined by Eq.~(\ref{Y}).
Inverting Eq.~(\ref{res}) in the reciprocal space, we can write
\begin{equation}
\delta \phi(\Gv \! + \! \qv,z,\omega) =
\sum\limits_{\Gv'} \int  \epsilon^{-1}_{\Gv\Gv'}(\qv,z,z',\omega) 
\times \left[ \delta \phi^\textit{ext}(\Gv' \! \! + \! \qv,z',\omega)+\delta \phi^I_{bare}(\Gv' \! + \! \qv,z',\omega) \right] d z'.
\end{equation}

Then we can write for the gradient of the effective potential 
\begin{equation}
\begin{split}
&\nabla \left. \delta \phi^\textit{eff}_{\alpha\0v}(\rv,\omega) \right|_{\rv=\bv_\alpha} = 
\sum\limits_{\Gv\Gv'} e^{i(\Gv+\qv)\cdot\bv_\alpha} \left[ i(\Gv+\qv) +\hat{\zv} \frac{\pa}{\pa z}  \right]
\int \left\{ \epsilon^{-1}_{\Gv\Gv'}(\qv,z',\omega) \phi^\textit{ext}(\Gv'+\qv,z',\omega) 
+ \frac{2\pi }{s_0} \times \right. \\ 
& \left. \sum\limits_{\beta} Z_\beta e^{-i(\Gv'  +  \qv)  \cdot  \bv_\beta}
\! \! \int \! \left[ \epsilon^{-1}_{\Gv\Gv'}(\qv,z', \omega) -\delta_{\Gv\Gv'} \delta(z-z')\right]
   \ev_\beta \cdot \Yv(z',\Gv'+\qv)  
   \right\}_{z=0} \! \! \! \! \! \! \! d z' \left . \! -  \! \! \! \! \! \! \sum\limits_{(\beta\Rv)\ne(\alpha\0v)} \! \! \! \! \! \! \! \!
   e^{i\qv\cdot\Rv} \nabla (\ev_\beta \! \cdot \! \nabla) \frac{Z_\beta}{|\rv \! - \! \bv_\beta \! - \! \Rv|} \right|_{\rv=\bv_\alpha} .
\end{split}
\end{equation}

We take use of the  static sum-rule \cite{Nazarov-05}
\begin{equation}
\nabla n_0(\rv)= -\int \chi(\rv,\rv',0) \nabla' v^\textit{ext}(\rv') \, d \rv'.
\end{equation}
Then
\begin{equation}
\int \frac{\nabla n_0(\rv')}{|\rv'-\rv|} d \rv' \! = \!
\lim_{\qv\rightarrow\0v}
\frac{2\pi}{s_0}  \! \! \sum\limits_{\beta\Gv\Gv'} \! Z_\beta  \int  \! \!
\left[ \epsilon^{-1}_{\Gv\Gv'}(\qv,z,z',0)    - \delta_{\Gv\Gv'} \delta(z \! - \! z') \right] 
\Yv(z',\Gv'+\qv) e^{i(\Gv+\qv)\cdot \rv} e^{-i(\Gv'+\qv) \cdot \bv_\beta} d z' .
\end{equation}

Finally, we have for the force acting on the $\alpha$-th nucleus
\begin{equation}
F_{\alpha i} \! = \! F^\textit{ext}_{\alpha i}(\qv,\omega) \! + \!  F^\textit{ei}_{\alpha i}(\qv,\omega)
\! + \! \! \sum\limits_{\beta k}\!  N_{\alpha i,\beta k} (\qv,\omega) e_{\beta k},
\label{force}
\end{equation}
\begin{equation}
N_{\alpha i,\beta k} (\qv,\omega)=P_{\alpha i,\beta k} (\qv,\omega)- \delta_{\alpha\beta}  \sum\limits_\gamma P_{\alpha i,\gamma k} (\0v,0),
\end{equation}
where
\begin{equation}
 F^\textit{ext}_{\alpha i}  =
-Z_\alpha \sum\limits_{\Gv} e^{i(\Gv+\qv)\cdot\bv_\alpha} 
\left. \left[ i(G_i+q_i) +\hat{z}_i \frac{\pa}{\pa z}  \right] \phi^\textit{ext}(\Gv+\qv,z,\omega) \right|_{z=0}, 
\end{equation}

\begin{equation}
F^\textit{ei}_{\alpha i} \!  = \!
  2\pi Z_\alpha \! \! \sum\limits_{\Gv\Gv'}  
\int \! 
Y_i(z,\Gv+\qv) \chi_{\Gv\Gv'}(\qv,z,z',\omega)  
e^{i(\Gv \! + \! \qv)\cdot\bv_\alpha}
\phi^\textit{ext}(\Gv' \! + \! \qv,z',\omega) d z d z',
\end{equation}
\begin{equation}
P_{\alpha i,\beta k} (\qv,\omega)=Q_{\alpha i,\beta k} (\qv,\omega)+S_{\alpha i,\beta k} (\qv),
\end{equation}
\begin{equation}
Q_{\alpha i,\beta k} (\qv,\omega) =
\frac{(2\pi)^2}{s_0} \sum\limits_{ \Gv\Gv'} Z_\alpha Z_\beta e^{i(\Gv+\qv)\cdot\bv_\alpha} e^{-i(\Gv'  +  \qv)  \cdot  \bv_\beta} 
 \int Y_i(z,\Gv+\qv)    \chi_{\Gv\Gv'}(\qv,z,z',\omega) 
  Y_k(z',\Gv'+\qv) d z d z',
\end{equation}
\begin{equation}
S_{\alpha i,\beta k} (\qv) \! = \!
\left. \! \! \sum\limits_\Rv \!
\left[1 \! - \! \delta_{\Rv\0v} \delta_{\beta\alpha} \right] e^{i\qv\cdot\Rv} \nabla_i \nabla_k
\frac{Z_\alpha Z_\beta}{|\rv \! - \! \bv_\beta \! - \! \Rv|} \right|_{\rv=\bv_\alpha} \! \! \! .
\end{equation}

Noting that within the static approximation ($\omega=0$) our theory reduces
to the conventional density-functional perturbation theory (DFPT) \cite{Giannozzi-05},
we can conveniently rewrite Eq.~(\ref{force}) as Eq.~(\ref{forceD}).

\end{document}